\documentclass[twocolumn]{aastex701}

\usepackage{enumitem}

\def\Rdisk{R_\mathrm{disk}}
\def\MJ{M_\mathrm{J}}

\begin{document}

\title{JWST/MIRI Imaging Search for Kinematically Detected Protoplanetary Candidates}

\author[0000-0001-7255-3251]{Gabriele Cugno}
\affiliation{\instUZH}
\email[]{gabriele.cugno@uzh.ch}

\author[0000-0002-7695-7605]{Myriam Benisty}
\affiliation{\InstMPIA}\email[]{}

\author[0000-0003-1534-5186]{Richard Teague}
\affiliation{\InstMIT}\email[]{}


\author[0000-0001-9353-2724]{Anthony Boccaletti}
\affiliation{\InstLESIA} \email[]{}
\author[]{Laurent Pueyo}
\affiliation{\InstSTScI}\email[]{}
\author[0000-0002-3191-8151]{Marshall Perrin}
\affiliation{\InstSTScI}\email[]{}
\author[0000-0002-2918-8479]{Mathilde Malin}
\affiliation{\InstJohnHop}\affiliation{\InstSTScI} \email[]{}
\author[0000-0001-8627-0404]{Julien Girard}
\affiliation{\InstSTScI}\email[]{}
\author[0000-0002-0101-8814]{Valentin Christiaens}
\affiliation{Université Paris-Saclay, Université Paris Cité, CEA, CNRS, AIM, F-91191 Gif-sur-Yvette, France}\email[]{}
\author[0000-0001-8718-3732]{Polychronis Patapis}
\affiliation{\InstETH}\email[]{}
\author[0000-0002-6964-8732]{Kellen Lawson}
\affiliation{Center for Space Sciences and Technology, University of Maryland, Baltimore County, 1000 Hilltop Circle, Baltimore, MD 21250, USA}
\affiliation{Astrophysics Science Division, NASA-GSFC, 8800 Greenbelt Rd, Greenbelt, MD 20771, USA}
\affiliation{Center for Research and Exploration in Space Science and Technology, NASA-GSFC, 8800 Greenbelt Rd, Greenbelt, MD 20771, USA}\email[]{}


\author[0000-0003-2253-2270]{Sean M. Andrews}
\affiliation{\InstCfA}\email[]{}

\author[0000-0001-7258-770X]{Jaehan Bae}
\affiliation{\InstFlorida}\email[]{}

\author[0000-0001-6378-7873]{Marcelo Barraza-Alfaro}
\affiliation{\InstMIT}
\affiliation{Niels Bohr International Academy, Niels Bohr Institute, Blegdamsvej 17, DK-2100 Copenhagen Ø, Denmark}\email[]{}

\author[]{Markus J. Bonse}
\affiliation{\InstESO}\affiliation{\InstETH}\email[]{}

\author[]{Margot Courtoux}
\affiliation{\InstLESIA}\email[]{}

\author[0000-0003-4689-2684]{Stefano Facchini}
\affiliation{\InstMilano}\email[]{}

\author[0000-0003-1117-9213]{Misato Fukagawa} 
\affiliation{\InstAIJ}\email[]{}

\author[]{Greta Guidi} 
\affiliation{\InstIPAGGrenoble}\email[]{}

\author[]{Ravit Helled} 
\affiliation{\instUZH}\email[]{}

\author[0000-0002-1493-300X]{Thomas Henning} 
\affiliation{\InstMPIA}\email[]{}

\author[0000-0001-6947-6072]{Jane Huang} 
\affiliation{\InstColumbia}\email[]{}

\author[0000-0003-2769-0438]{Jens Kammerer} 
\affiliation{\InstESO}\email[]{}

\author[0000-0003-1413-1776]{Charles J.\ Law}
\affiliation{Minnesota Institute for Astrophysics, University of Minnesota, 116 Church St. SE, Minneapolis, MN 55455}
\email{}

\author[0000-0002-2357-7692]{Giuseppe Lodato} 
\affiliation{\InstMilano}\email[]{}

\author[0000-0002-7607-719X]{Feng Long}
\affiliation{\InstKavli}\email[]{}

\author[0000-0002-1637-7393]{François M\'{e}nard} 
\affiliation{\InstIPAGGrenoble}\email[]{}

\author[0000-0003-1227-3084]{Michael Meyer}
\affiliation{Department of Astronomy, University of Michigan, 1085 S. University Ave., Ann Arbor, MI 48109, USA}
\email[]{} 

\author[0000-0002-4716-4235]{Daniel J. Price} 
\affiliation{\InstMonash}\email[]{}

\author[0000-0003-1817-6576]{Christian Rab} 
\affiliation{\InstMPE}\affiliation{\InstLMU}\email[]{}

\author[]{Isabel Rebollido Vazquez} 
\affiliation{\InstESA}\email[]{}

\author[0000-0001-5907-5179]{Christophe Pinte}
\affiliation{\InstIPAGGrenoble}\email[]{}

\author[0000-0003-4853-5736]{Giovanni Rosotti} 
\affiliation{\InstMilano}\email[]{}

\author[0000-0003-3829-7412]{Sascha Quanz} 
\affiliation{\InstETH}\email[]{}

\author[0000-0002-5991-8073]{Anibal Sierra} 
\affiliation{\InstUNAM}\email[]{}

\author[0000-0002-0491-143X]{Jochen Stadler} 
\affiliation{\InstESO}\email[]{}

\author[0000-0001-9524-3408]{Lucas M. Stapper} 
\affiliation{\InstMPIA}\email[]{}

\author[0000-0002-5823-3072]{Tomas Stolker} 
\affiliation{\InstLeiden}\email[]{}

\author[0000-0003-1526-7587]{David J. Wilner} 
\affiliation{\InstCfA}\email[]{}

\author[0000-0002-7501-9801]{Andrew J. Winter}
\affiliation{\InstQMUL}\email[]{}


\newcommand{\InstMPIA}{Max-Planck Institute for Astronomy (MPIA), Königstuhl 17, 69117 Heidelberg, Germany}
\newcommand{\instUZH}{Department of Astrophysics, University of Z\"urich, Winterthurerstrasse 190, 8057 Z\"urich, Switzerland}
\newcommand{\InstMPE}{Max-Planck-Institut für extraterrestrische Physik, Giessenbachstrasse 1, 85748 Garching, Germany}
\newcommand{\InstMIT}{Department of Earth, Atmospheric, and Planetary Sciences, Massachusetts Institute of Technology, Cambridge, MA 02139, USA}
\newcommand{\InstMilano}{Dipartimento di Fisica, Università degli Studi di Milano, Via Celoria 16, 20133 Milano, Italy}
\newcommand{\InstNAOJ}{National Astronomical Observatory of Japan, Osawa 2-21-1, Mitaka, Tokyo 181-8588, Japan}
\newcommand{\InstAIJ}{Astronomical Institute, Graduate School of Science, Tohoku University, 6-3 Aoba, Aramaki, Aoba-ku, Sendai, Miyagi 980-8578 Japan}
\newcommand{\InstIPAGGrenoble}{Univ. Grenoble Alpes, CNRS, IPAG, 38000 Grenoble, France}
\newcommand{\InstMonash}{School of Physics and Astronomy, Monash University, Clayton VIC 3800, Australia}
 \newcommand{\InstCfA}{Center for Astrophysics | Harvard \& Smithsonian, Cambridge, MA 02138, USA}
\newcommand{\InstFlorida}{Department of Astronomy, University of Florida, Gainesville, FL 32611, USA}
 \newcommand{\InstColumbia}{Department of Astronomy, Columbia University, 538 W. 120th Street, Pupin Hall, New York, NY, USA}
\newcommand{\InstLeiden}{Leiden Observatory, Leiden University, P.O. Box 9513, NL-2300 RA Leiden, The Netherlands}
\newcommand{\InstESO}{European Southern Observatory, Karl-Schwarzschild-Str. 2, D-85748 Garching bei München, Germany}
\newcommand{\InstVirginia}{Department of Astronomy, University of Virginia, Charlottesville, VA 22904, USA}
\newcommand{\InstUNAM}{Universidad Nacional Autónoma de México. Instituto de Astronomía. A.P. 70-264, 04510. Ciudad de México, México}
\newcommand{\InstQMUL}{Astronomy Unit, School of Physics and Astronomy, Queen Mary University of London, London E1 4NS, UK}
\newcommand{\InstSTScI}{Space Telescope Science Institute, 3700 San Martin Drive, Baltimore, MD 21218, USA}
\newcommand{\InstLESIA}{LESIA, Observatoire de Paris, Universit\'e PSL, CNRS, Sorbonne Université, Univ. Paris Diderot, Sorbonne Paris Cité, 5 place Jules
Janssen, 92195 Meudon, France}
\newcommand{\InstJohnHop}{Department of Physics \& Astronomy, Johns Hopkins University, 3400 N. Charles Street, Baltimore, MD 21218, USA}
\newcommand{\InstKavli}{Kavli Institute for Astronomy and Astrophysics, Peking University, Beijing 100871, China}
\newcommand{\InstLMU}{University Observatory, Faculty of Physics, Ludwig-Maximilians-Universität München, Scheinerstr. 1, 81679 Munich, Germany}
\newcommand{\InstESA}{European Space Agency(ESA),European Space Astronomy Centre(ESAC), Camino Bajo del Castillo s/n,28692, Villanueva de la Canada, Madrid, Spain}
\newcommand{\InstETH}{ETH Zürich, Institute for Particle Physics and Astrophysics, Wolfgang-Pauli-Str. 27, 8093 Zürich, Switzerland}

\begin{abstract}
Kinematic perturbations observed with ALMA in CO line emission provide evidence for a population of embedded giant protoplanets shaping the structure of protoplanetary disks. 
We present JWST/MIRI F1140C ($\lambda=11.3~\mu$m) coronagraphic observations of five protoplanetary disks, HD 163296, RXJ1615.3–3255, RXJ1842.9–3532, SY Cha, and LkCa 15, with the goal of directly detecting candidate protoplanets orbiting at $\gtrsim70$~au previously inferred from gas kinematics. The data were analyzed using a bespoke methodology that combines reference PSF subtraction with forward modeling of partially resolved inner disk emission, which otherwise dominates the diffraction pattern in the images. This approach improves the sensitivity to young companions at small separations. No point source consistent with an embedded protoplanet is detected in any of the systems. Instead, in three systems we detect extended emission at $11.3~\mu$m tracing the outer disk out to radii comparable to those probed by CO. Injection tests indicate upper mass limits of roughly $3–20~\MJ$ at separations of a few hundred au, assuming no additional thermal contribution from circumplanetary environment. Even with space-based observations, these limits remain mostly above the $\sim1-5~\MJ$ masses inferred from disk kinematics, largely due to the limitations imposed by emission (and/or scattered light) contributions from both the inner and outer disk. These observations highlight the challenges of observing protoplanets embedded in their forming environment at large separation with JWST/MIRI. Lessons learned can inform future studies with the Extremely Large Telescope, which will probe separations where the occurrence rate of gas giants is expected to be higher. 

\end{abstract}

\keywords{\uat{High contrast techniques}{2369} --- \uat{Exoplanet formation}{492} --- \uat{Protoplanetary disks}{1300}}


\section{Introduction} 

High-resolution sub-millimeter and infrared imaging campaigns have shown that substructures (e.g., spirals, gaps and rings), appear to be ubiquitous in the gas and dust distributions of large and bright protoplanetary disks \citep{Andrews2020, Oberg2021}. While other possible explanations exist \citep[e.g., ][]{Flock2015}, some of these features are often thought to be the result of planet--disk interactions, and their ubiquity indicates that a large population of yet-undetected giant planets at large separation might be sculpting the structure of protoplanetary disks \citep[e.g.,][]{Dipierro2015, Perez2016, Huang2018, Bae2023, Ruzza2025}. This has motivated extensive direct-imaging efforts from both ground- and space-based facilities, spanning optical to infrared wavelengths. With the exception of two systems with unambiguous detections, PDS70 and WISPIT2 \citep[e.g.,][]{Keppler2018, van_Capelleveen2025}, these surveys have however resulted in non-detections \citep[e.g.,][]{Cugno2019, Asensio-Torres2021, Huelamo2022, Ren2023, Cugno2023}, with current sensitivity limits reaching a few Jupiter masses for ground-based observations \citep{Ren2023}, and sub-Jupiter masses for space-based observations outside of disks with JWST/NIRCam \citep{Cugno2024_NIRCam}, at separations $\gtrsim2\arcsec{}$. The paucity of detections may be attributed to significant extinction from the circumstellar disk surrounding the forming planet that can obscure its emission \citep{Alarcon2024, Cugno2025}.

While planets may remain hidden by the disk at infrared wavelengths, they are expected to leave clear signatures in the velocity field of their host circumstellar disk that can be imaged using CO emission lines with ALMA \citep{Teague2018a,Pinte2019,Pinte2022,Izquierdo2022}. As a consequence, local departures from purely Keplerian rotation have been suggested as a diagnostic for detecting embedded companions or planets \citep{Pinte2018b}. Such velocity deviations (with magnitudes up to $\sim$10\% of the background Keplerian rotation) were recently detected, manifesting as `kinks' (thunderbolt-shaped perturbations) in channel maps \citep[e.g.,][]{Pinte2018,Pinte2019} and `Doppler flips' \citep[e.g.,][]{Casassus_Perez_2019} in residual velocity maps after subtraction of the background Keplerian rotation. These perturbations can be associated with spiral wakes \citep{Perez2015, Bollati2021, Calcino2022}  
and their amplitude strongly depends on the mass of the embedded planet. As such, kinematical perturbations provide a complementary planet-hunting technique to direct imaging. 

\begin{table*}[t!]
\centering
\caption{Summary of observations.}
\def\arraystretch{1.25}
\begin{tabular}{llllllllllll}\hline
Target & Prog. ID & Filter & $\lambda_\mathrm{pivot}$ & W$_\mathrm{eff}$ & Readout & Subarray & $N_\mathrm{gr}$ & $N_\mathrm{int}$ & $N_\mathrm{dither}$ & $N_\mathrm{roll}$ & $t_\mathrm{tot}$  \\
& & & ($\mu$m)  & ($\mu$m) & & & & & & & (s)  \\ \hline
HD163296 & 2153 & F1140C  & 11.3 & 0.8 & FASTR1 & MASK1140 & 200 & 172 & 1 & 2\tablenotemark{a} & 16570  \\
J1615 & 3254 & F1140C  & 11.3 & 0.8 & FASTR1 & MASK1140 & 500 & 30 & 1 & 1 & 3602      \\
J1842 & 3254 & F1140C  & 11.3 & 0.8 & FASTR1 & MASK1140 & 500 & 40 & 1 & 1 & 4802  \\
SY Cha & 3254 & F1140C  & 11.3 & 0.8 & FASTR1 & MASK1140 & 500 & 40 & 1 & 1 & 4802  \\ 
LkCa15    & 3254 & F1140C  & 11.3 & 0.8 & FASTR1 & MASK1140 & 500  & 26 & 1 & 1 & 3121   \\ \hline
\end{tabular}\\\vspace{0.2cm}
\tablenotemark{a} Both visits where taken with the same spacecraft orientation, resulting in the same roll angle. 
\label{tab:observations}
\end{table*}

Kinematic signatures of possible embedded planets were first showcased on the multi-ringed disk around the Herbig Ae star HD 163296, one of the objects studied in this paper. \citet{Pinte2018} identified in ALMA CO observations a localized, non-Keplerian signature at $\sim$260 au, which they interpreted as the dynamical imprint of a $\sim2\MJ$ planet. This interpretation was further supported by \cite{Calcino2022}, who reported extended kinematic perturbations consistent with the planetary wake launched by this candidate. Using accurate azimuthally averaged rotation curves, rather than localized perturbations, \citet{Teague2018a} identified axisymmetric deviations from Keplerian rotation that trace changes in the radial pressure gradient and can be attributed to gaps induced by planet–disk interactions. Comparisons with hydrodynamical models showed that the observed velocity structure is best reproduced by the presence of two Jupiter-mass planets orbiting at approximately 83 au and 137 au. More recently, \citet{Izquierdo2022} applied a channel-map modeling approach to the same system and, in addition to the planet candidate previously inferred at $\sim$260 au, revealed a new $\sim1~\MJ$ planet candidate at $\sim$94 au, likely tracing the same feature previously associated with the candidate at 83 au. This candidate's location coincides with a dust gap observed in the continuum emission, where evidence for a candidate circumplanetary disk (CPD) has been reported \citep{Izquierdo2026}. Additional support for a planet-driven origin of this gap comes from CO isotopologue modeling, which identifies a gas-density gap at the same location \citep{Zhang2021}, and from multiline CS observations showing locally reduced gas density consistent with a Jupiter-mass perturber \citep{Law2025}.


Building on these works, the ALMA Large Program exoALMA \citep{Teague2025} observed 15 disks with high angular and spectral resolution observations. Most of these disks show ubiquitous velocity substructures across their spatial extents \citep{Stadler2025, Fukagawa2026} that can be attributed to a combination of physical mechanisms, such as hydrodynamical instabilities \citep[][]{Barraza2025, Benisty2026} 
and planet-disk interactions \citep{Izquierdo2026b}. In this sample, \citet{Pinte2025} identified a sub-sample of six disks with significant
, localized, velocity deviations that provide evidence for embedded $1-5~\MJ$ protoplanets at orbital separations between 80 and 310 au. 

In this paper, we present the first JWST/MIRI coronagraphic search for embedded protoplanets identified through gas kinematics. We target five highly structured disks with rings and gaps that have kinematically-detected planet candidates: 
HD163296 \citep{Pinte2018b, Izquierdo2022} from Program GO 2153, and SY~Cha, RXJ1842.9-3532 (J1842), RXJ1615.3-3255 (J1615), LkCa 15 \citep{Pinte2025} from Program GO 3254. Among the six targets presented in \citet{Pinte2025}, one (AA Tau) is too highly inclined, while another (HD 143006) has not undergone detailed modeling comparable to that available for the other four targets. We therefore limited our observations to the remaining four targets. 
In those systems, the observed velocity signatures point to candidate companions in the giant-planet regime, at wide separations that are in principle accessible to direct imaging. Leveraging the unique sensitivity enabled by JWST, we aim to directly image the companion candidates responsible for the observed velocity signatures. A direct confirmation of a protoplanet would provide a validation of gas kinematics as a method for detecting forming planets. The paper is structured as follows: 
we first present the observations (Sect.\,\ref{sec:obs}) and the method employed to reduce the data and model the inner disk contribution (Sect.\,\ref{sec:methods}).  Section \ref{sec:results} presents the final images and the detection limits extracted from them. In  Sect.\,\ref{sec:discussion}, we discuss our findings and conclude in Sect.\,\ref{sec:conclusion}.


\section{Observations}
\label{sec:obs}

We observed five protoplanetary disks with JWST/MIRI \citep{Wright2015} using its mid-infrared coronagraphic mode, which employs Four-Quadrant Phase Masks (FQPMs, \citealt{Boccaletti2015, Boccaletti2022}) to achieve high contrast at small angular separations. Out of the 3 filters available in coronagraphy, all targets were observed with the F1140C filter ($\lambda_c = 11.3~\mu$m, $\Delta\lambda=0.8~\mu$m). This filter provides better angular resolution than F1550C ($\lambda=15.5~\mu$m) and avoids the stronger silicate features in F1065C ($\lambda=10.65~\mu$m) that would otherwise increase the flux from the central source and contrast requirements.

Each sequence followed the standard MIRI coronagraphic pattern: science target, science background, reference background, and reference star observations. The number of groups, integrations, and total exposure times for each target are given in Table \ref{tab:observations}. Reference star observations have been executed using the small grid dither (SGD) pattern to increase diversity in the PSF feature, as it is standard procedure with JWST. All observations were obtained at a unique spacecraft roll angle. For HD163296, a mistake in the preparation of the Astronomer's Proposal Tool (APT) produced two visits with identical roll angles, providing no additional diversity for PSF subtraction.

PSF subtraction was performed using Reference Differential Imaging (RDI). Reference stars for these programs were chosen to match brightness and color of the target star in the mid-IR (for Program Identifier, PID, 2153) and through the JMMC \texttt{SearchCal}\footnote{Available at https://www.jmmc.fr/searchcal} tool (for PID3254). Two of these reference observations exhibited significant issues, namely the presence of resolved emission (PID2153, associated with HD163296) and a failed Target Acquisition (TA) procedure (PID3254, obs 12, associated with J1615). The other reference observations obtained as part of PID3254 resulted in comparable detector count levels in the MIRI pixels illuminated by the stellar PSF. However, because the reference stars were selected to be slightly brighter than the science targets, the shorter integration times resulted in comparatively higher background noise that would limit the sensitivity of the science data when performing PSF subtraction. In addition, for the remaining targets with nominal reference observations, partially resolved inner disk emission alters the PSF morphology, reducing the effectiveness of PSF subtraction \citep{Boccaletti2024}. To mitigate all these effects, we expanded our reference star library beyond our own observations to increase the diversity of PSF structures and to enhance our ability to subtract stellar residuals \citep[also see][]{Malin2025}. We included all F1140C reference datasets acquired before May 2025, drawing from JWST Programs 1194, 1277, 1386, 1668, 2538, 3662, and 5037. 

\begin{figure*}[t]
    \centering
    \includegraphics[width=0.95\linewidth]{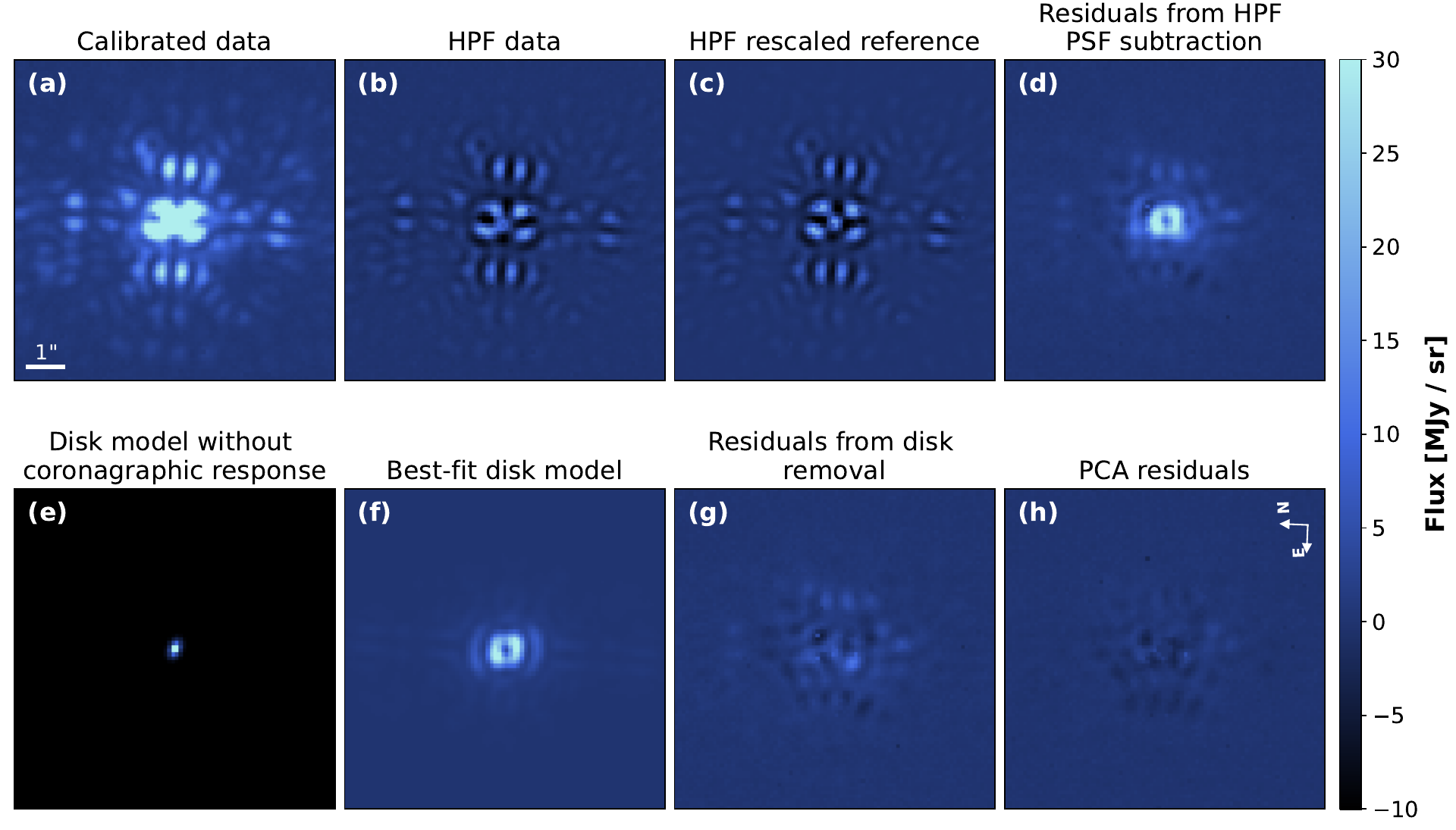}
    \caption{Step by step illustration of the removal of the unresolved PSF and  resolved inner disk signals to search for embedded protoplanets. The image shows the processes applied to the LkCa15 dataset as an example. The top row focuses on the PSF removal using high-pass filtered (HPF) images (Sect.~\ref{sec:initPSF}), while the bottom row demonstrates the steps presented in Sects.~\ref{sec:innerdisk} and~\ref{sec:PCA}, modeling and removing the inner disk component. With the exception of panel (e), which is a normalized model shown at the pixel resolution of the MIRI detector, the color scale is the same for all the other panels. Images have not been derotated and are shown as they appear on the detector. Panel (h) shows the orientation with respect to North and East.}
    \label{fig:methods}
\end{figure*}

\section{Data reduction}\label{sec:methods}

\subsection{Pre-processing}
\label{sec:preproc}
All science and reference data were processed uniformly using the {\tt jwst} pipeline, running Stages 1 and 2 starting from the {\tt uncal.fits} files. Following MIRI coronagraphy recommendations \citep{Carter2023}, we skipped the dark-current and flat-fielding corrections step and adopted a jump-detection threshold of 8. For each exposure, the first frame was discarded, and remaining bad pixels were manually identified and replaced with the mean of the surrounding pixels. Background observations were subtracted from their corresponding science and reference frames to remove thermal background and glowsticks \citep[see][]{Boccaletti2022}. The background observations of HD163296 were heavily contaminated by background stars, and we manually masked sources in both realizations before combining them into a final background image. Finally, all science and reference exposures were median-combined, yielding one image per visit/dither for each target and a set of 9 (or 5, depending on the program) reference-star frames. 

\subsection{Removal of the central emission (star+inner disk)}
As an example, the combined image for LkCa15 resulting from the pre-processing is shown in Fig.~\ref{fig:methods}a. For all 5 objects, bright disk emission, partially resolved by MIRI (hereafter referred to as the `inner disk') contributes significantly to the diffraction pattern seen in the combined images. This inner disk that contributes to the Mid-IR images can be several 10s of AU in size, and should not be confused with what is normally referred to as inner disk, namely the few AU region closest to the star. In addition, for some objects, we detect extended signal at large separations, from the outer disk regions (see Sect.~\ref{sec:disks}). 

The presence of  inner disk component (at the MIRI spatial resolution of $\approx0\farcs36$) makes standard PCA-based PSF subtraction inefficient \citep[e.g.,][]{Carter2023}, as the algorithm tends to over-estimate the contribution from the stellar diffraction and thus over-subtract the stellar PSF. This issue is known from previous works on debris disks, where the stellar PSF was instead removed using a scaled version of a reference PSF \citep[e.g.,][]{Malin2024} or a linear combination optimized at separations beyond the disk emission \citep{Boccaletti2024}. Protoplanetary disks however, are brighter and more extended than debris disks, and the resulting diffraction pattern affects significantly larger separations than in systems such as HR8799 or HD95086 \citep{Boccaletti2024, Malin2024}. To mitigate these issues, we 
first removed the main PSF contribution using high-pass filtered PSFs (Section~\ref{sec:initPSF}), then modeled the inner disk contribution (Sect.~\ref{sec:innerdisk}), and performed a final step based on Principal Component Analysis (PCA) when applicable (Section~\ref{sec:PCA}).

\begin{table*}[t!]
\centering
\caption{Information on the PSF subtraction process applied to each target and results from the inner disk fits.}
\def\arraystretch{1.25}
\begin{tabular}{lllllllll}\hline
 & \multicolumn{2}{c}{PSF step \#1} & & \multicolumn{4}{c}{Inner disk modeling} & PSF step \#2 \\
Target   & Technique\tablenotemark{a} & Ref PID, visit & & $F$ & \multicolumn{2}{c}{$\Rdisk$} & $b$ & Technique\tablenotemark{a} \\
         &        &                     & &  & (au) & ($''$) & (MJy/sr) &\\\hline
HD163296 & SR & 1668, 14 & & $74055.7^{+4705.3}_{-4811.4}$ 
         & $137.6^{+41.4}_{-43.5}$ & $1.36^{+0.41}_{-0.43}$ 
         & $551.7^{+82.0}_{-82.6}$ & SR \\

J1615    & LC & 2538, 4 & & $51.5^{+3.0}_{-3.8}$ 
         & $85.7^{+2.4}_{-4.1}$ & $0.55^{+0.02}_{-0.03}$ 
         & $2.50^{+0.02}_{-0.02}$ & LC \\

J1842    & LC & 2538, 114 & & $372.0^{+39.3}_{-26.5}$ 
         & $76.6^{+7.9}_{-9.0}$ & $0.51^{+0.05}_{-0.06}$ 
         & $1.6^{+0.2}_{-0.2}$ & LC \\

SY Cha   & SR & 5037, 3 & & $309.3^{+31.1}_{-19.9}$ 
         & $139.1^{+28.2}_{-30.9}$ & $0.76^{+0.15}_{-0.17}$ 
         & $0.4^{+0.2}_{-0.2}$ & PCA \\ 

LkCa15   & SR & 3662, 9 & & $846.5^{+122.7}_{-85.1}$ 
         & $71.2^{+8.8}_{-9.0}$ & $0.45^{+0.06}_{-0.06}$ 
         & $1.6^{+0.3}_{-0.3}$ & PCA \\ \hline
\end{tabular}\\\vspace{0.2cm}
\tablenotemark{a} SR = Single Reference; LC = Linear Combination; 
PCA = Principal Component Analysis. 
\label{tab:results}
\end{table*}

\subsubsection{Initial PSF subtraction}
\label{sec:initPSF}

To remove the extended disk signal, we first high-pass filtered both the science and reference images using a Gaussian kernel with FWHM = 2.8 pixels (approximately the MIRI PSF size at $11.3~\mu$m), verifying that variations in the Gaussian FWHM have a negligible impact on the final residuals. While these disks exhibit narrow ($\sim0\farcs1$) substructures at near-infrared wavelengths \citep[e.g.][]{Ren2023}, the observed morphology of such features is broadened to approximately the PSF scale. As this filtering removes spatial structures larger than the kernel scale, it suppresses most of the extended disk emission while preserving the stellar diffraction pattern (see Fig.~\ref{fig:methods}b). We emphasize that these images are used only to determine the optimal reference scaling; the resulting coefficients are subsequently applied to the original, unfiltered reference images.

To remove the PSF we undertook two approaches. First, similar to \cite{Malin2024}, for each exposure we iteratively determined the optimal scaling factor between the high-pass-filtered reference and science images (Fig.~\ref{fig:methods}c shows the results for the first PSF dither in the observation 9 from PID3662). Once this factor was found, we applied it to the original (unfiltered) reference frame and subtracted the scaled reference from the science image. The scaling factor was obtained by minimizing the sum of the squared values of the residual pixel intensities within an annulus between 10 and 25 pixels in radius ($1\farcs1-2\farcs7$), a region where the stellar diffraction pattern remains strong but the contribution from the partially resolved inner disk is reduced. As we expect some positive signal in the image due to potential diffraction from the bright disk, we penalize negative pixels by a factor 10 in the cost function as to reduce the risk of oversubtraction. Because this optimization is performed for each reference image, it yields between $5-9$ residual frames per reference star (depending on the choice SGD pattern used). In practice though, for each reference star there is one image that produces better results due to the more similar position of the star behind the coronagraph as in the science data. This best-performance frame was used to obtain Fig.~\ref{fig:methods}d, and in the second column of Tab.~\ref{tab:results} is indicated as SR (Single Reference).

The second approach to remove the stellar diffraction followed the {\tt amoeba} method described in \cite{Boccaletti2024}. Here, the stellar diffraction is reproduced by the linear combination of each set of 5 or 9 reference dithers. We applied the same mask as above to avoid the central region dominated by the resolved disk signal, which means that we have limited control on how correctly the linear combination of PSFs reproduces the stellar signal within the inner $1\farcs0$. In the second column of Tab.~\ref{tab:results} this method is indicated as LC (Linear Combination).

Both approaches were applied to each target, and the most effective PSF subtraction was decided upon visual inspection of the residuals, minimizing the presence of negative pixels due to oversubtraction and ensuring, when possible, regular and symmetric signals of the inner disk component. This implies the PSF signal is well-subtracted and there was no oversubtraction on either sides of the star. For LkCa 15, the resulting optimized residual image is shown in Fig.~\ref{fig:methods}d.
The reference used for this process for each target is reported in the second column of Tab.~\ref{tab:results}. While this procedure removes most of the stellar diffracted light, residual structures associated with the partially resolved inner disk remain and dominate the image in the innermost regions.

\begin{figure*}[t]
    \centering
    \includegraphics[width=0.99\linewidth]{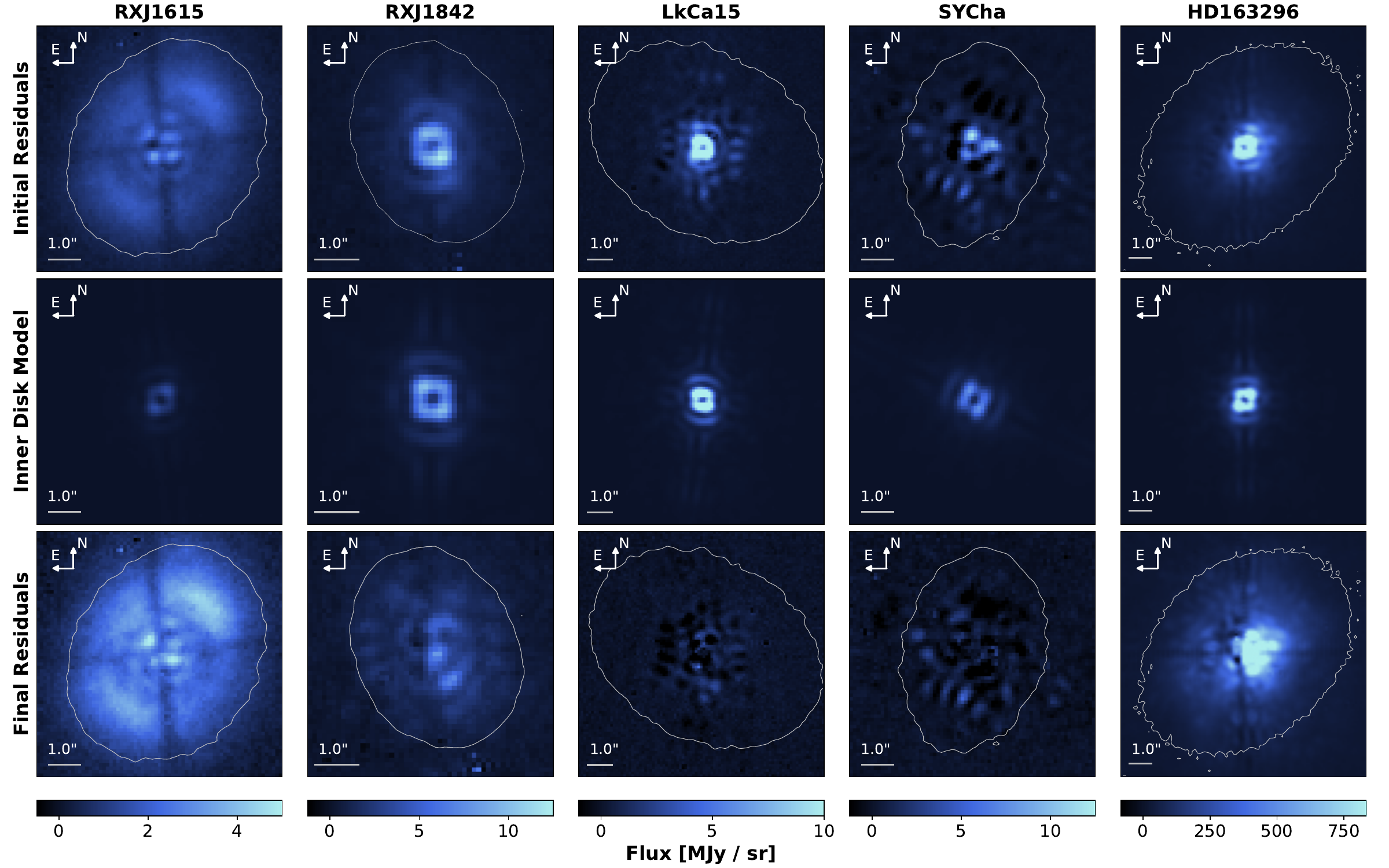}
    \caption{Residual gallery of our sample. {\it Top row:} Residuals after the initial PSF subtraction performed either with a single or a linear combination of reference PSFs. The white contour represents the outer edges of the 0th moment $^{12}$CO $J=3-2$ maps.
    {\it Middle row:} models of the inner disk emission obtained from the median values for each parameters. {\it Bottom row:} Residuals after removal of the inner disk signals. The asymmetry seen in J1842 and HD163296 is likely an artifact of the PSF subtraction algorithm. The limits of the colorscale have been halved with respect to the previous plots. The cross-shaped feature visible most prominently in RXJ1615 traces the transition regions of the MIRI four-quadrant phase mask and is therefore instrumental rather than astrophysical}
    \label{fig:gallery}
\end{figure*}

\subsubsection{Modeling the disk signal}
\label{sec:innerdisk}

To further remove the remaining contribution of the inner disk, we model the remaining emission by constructing a parametric disk model that reproduces the morphology of the disk, similar to the Model-Constrained RDI (MCRDI) technique presented in \cite{Lawson2022} and widely used on JWST/NIRCam disk images \citep[e.g.,][]{Lawson2023}. Each disk model is generated on a 100×100 pixel grid using the inclination and position angle of the outer disk reported in the literature \citep{Huang2018, Curone2025}. The free parameters are the disk radius $\Rdisk$, an overall multiplicative flux factor $F$ and a uniform background component $b$ (representing both detector background and any large-scale extended emission, as observed in systems like HD163296). We emphasize that the fitted $\Rdisk$ should be regarded as an effective characteristic extent of the emission required to reproduce the observed diffraction pattern, rather than as a precise measurement of the physical radius of the mid-infrared emitting region. The intrinsic disk brightness profile scales as $1/r^2$. Before projecting onto the model grid, the disk is rotated to match the on-detector orientation for each target. The resulting model for LkCa15 is shown in Fig.~\ref{fig:methods}e. It is striking that the inner disk contribution responsible for the artifacts in Fig.~\ref{fig:methods}d is less than $4$ pixels ($\lesssim0\farcs44$) in radius (see Tab.~\ref{tab:results}) in terms of angular extent and gets diffracted by the FQPM coronagraph. 

The disk image results from the linear combination of precomputed MIRI point sources at different detector locations, weighted by the flux distribution of the disk model at each corresponding position. To generate the set of PSF models at multiple locations, we simulated MIRI coronagraphic PSFs across the detector using {\tt stpsf}, at half-pixel sampling in both detector dimensions. Each PSF was computed without using Optical Path Difference (OPD) maps as the wavefront-error information at the coronagraphic location is only extrapolated and not directly measured, leaving significant and untested uncertainties. Additionally, we adopted a pupil shear of 2\% in both $x$ and $y$ directions.

\begin{figure*}[t]
    \centering
    \includegraphics[width=0.89\linewidth]{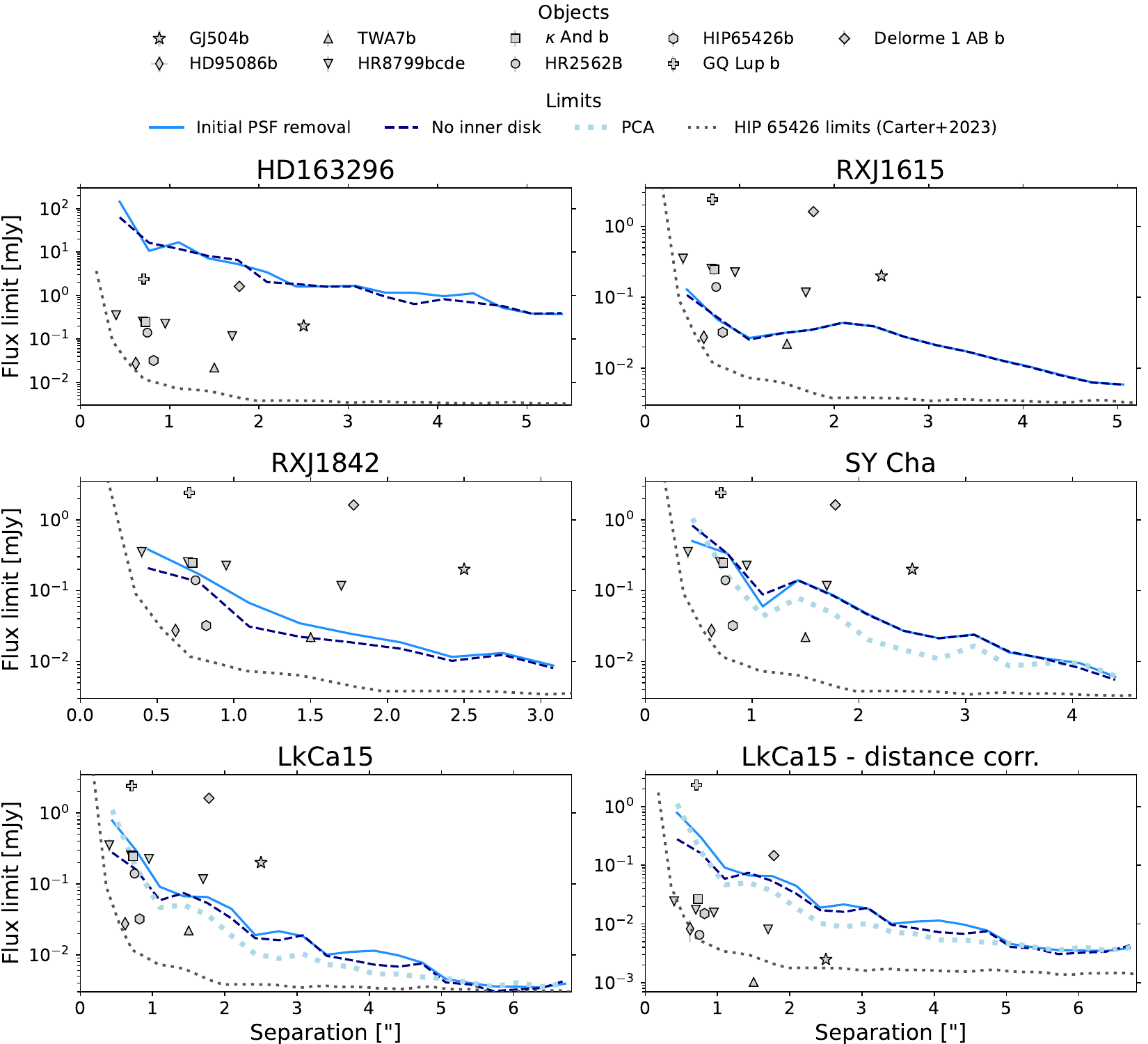}
    \caption{S/N=5 detection limits for our MIRI data, as measured after the initial PSF subtraction (Sect.~\ref{sec:initPSF}, solid blue line) and after the modeling and removal of the inner disk (Sect.~\ref{sec:innerdisk}, dashed line) and for SY Cha and LkCa15 after PCA-based PSF subtraction (Sect.~\ref{sec:PCA}, dotted line). The companions with measured F1140C fluxes from the literature are also reported \citep{Carter2023, Cugno2024, Malin2024, Godoy2024, Boccaletti2024, Lagrange2025, Malin2025_delorme, Godoy2025}, while the grey dashed line shows the detection limits reported by \cite{Carter2023} in the F1140C filter for HIP65426. The bottom right panel reports the same detection limits for LkCa15 reported in the bottom left panel, but this time the fluxes of the known companions and the HIP65426 limits are scaled to the distance of LkCa15.  }
    \label{fig:limits}
\end{figure*}

We explored the parameter space using {\tt pymultinest} \citep{Feroz2009,Buchner2016}, adopting uniform priors for $\Rdisk$ (between $0$ and $200$~au), $b$ (between $0$ and $1500$ MJy/sr) and $F$ (between 0 and $10^6$). The log-likelihood follows \citet{Christiaens2021} and is defined as
\begin{equation}
    L = -0.5 \sum_i \frac{R_i^2}{\sigma^2}
\end{equation}
where $R_i$ is the model residual at pixel $i$ within an annular region between 2 and 15 pixels from the center, and $\sigma$ is the standard deviation measured in the same area. The model resulting from the median of the posterior distributions for the three parameters is shown in Fig.~\ref{fig:methods}f. Subtracting this model from the initial residuals in Fig.~\ref{fig:methods}d yields the residual map displayed in Fig.~\ref{fig:methods}g. A comparison between panels~\ref{fig:methods}d and \ref{fig:methods}g demonstrates the dramatic improvement in the detectability of protoplanets in the innermost disk region when the disk signal is modeled and removed, despite residuals remaining dominated by speckle noise. Table~\ref{tab:results} provides, for each target, the median and $1\sigma$ uncertainty (calculated as the 16th/84th percentile of the posterior) resulting for each parameter from the inner disk modeling.

We note that several of these disks exhibit rings, gaps, and other substructures in higher-angular-resolution Near-IR observations \citep[e.g.][]{Ren2023}. We do not include these features explicitly in our inner disk model. We also assume an intrinsically axisymmetric brightness distribution centered on the star, neglecting potential azimuthal brightness variations, for example from anisotropic scattering, as well as apparent offsets that could arise if the 11.3~$\mu$m emission originates from an elevated disk surface. This is motivated by several factors: (i) the morphology at 11.3~$\mu$m is unknown, including its emission surface; (ii) most of the reported Near-IR substructures have characteristic widths smaller than the F1140C angular resolution and would therefore be strongly smoothed in the MIRI images; (iii) uncertainties in the current modeling and characterization of the MIRI coronagraphs can introduce systematics that could bias the inference of substructures and azimuthal asymmetries. Consequently, our goal is not to reproduce the detailed physical morphology of the inner disk, but to capture the dominant spatial extent and brightness distribution responsible for the additional diffraction pattern. A more detailed characterization of the instrumental response to resolved emission will enable more complex disk models and is left for future work (Courtoux et al., in prep.).

\subsubsection{PCA PSF subtraction}
\label{sec:PCA}

After determining the best-fit disk model, we return to the calibrated (non–high-pass-filtered and non-PSF-subtracted) data and subtract the forward-modeled disk contribution. This step removes the extended inner-disk emission. For LkCa15 and SY Cha, this leaves images that contain only the stellar PSF and any potential unresolved sources, such as protoplanets emitting in the mid-infrared. In these cases, to remove the remaining stellar PSF, we apply Principal Component Analysis (PCA, \citealt{Soummer2012, AmaraQuanz2012}), constructing a PSF basis from all available F1140C reference-star observations. The science images are then projected onto this basis and the reconstructed PSF is subtracted. For LkCa15, the residual image obtained after subtracting 15 components is shown in Fig.~\ref{fig:methods}h. This method has been used extensively for point source detection in NIRCam coronagraphy of circumstellar disks \citep[e.g.,][]{Lawson2022, Lawson2023}

In the other cases, extended signals tracing the entirety of the disks remain in the images, making PCA prone to over-estimate the intensity of the stellar diffraction in an attempt to get rid of the disk signal as well, resulting in strong over-subtraction. In those cases, we found that repeating the initial PSF subtraction step after removal of the inner resolved disk signal was enough to provide a cleaner residual. The last column in Table~\ref{tab:results} details the method used for each target for the final PSF subtraction.

\section{Results}\label{sec:results}

Figure~\ref{fig:gallery} shows, for all targets, the residuals obtained with the procedure described in Sect.\,\ref{sec:methods}, with the initial residuals in the top row, the inner disk model in the middle row and the final residuals in the bottom row. The residual maps reveal no clear point source that could be associated with embedded protoplanets. 

\begin{figure*}
    \centering
    \includegraphics[width=0.8\linewidth]{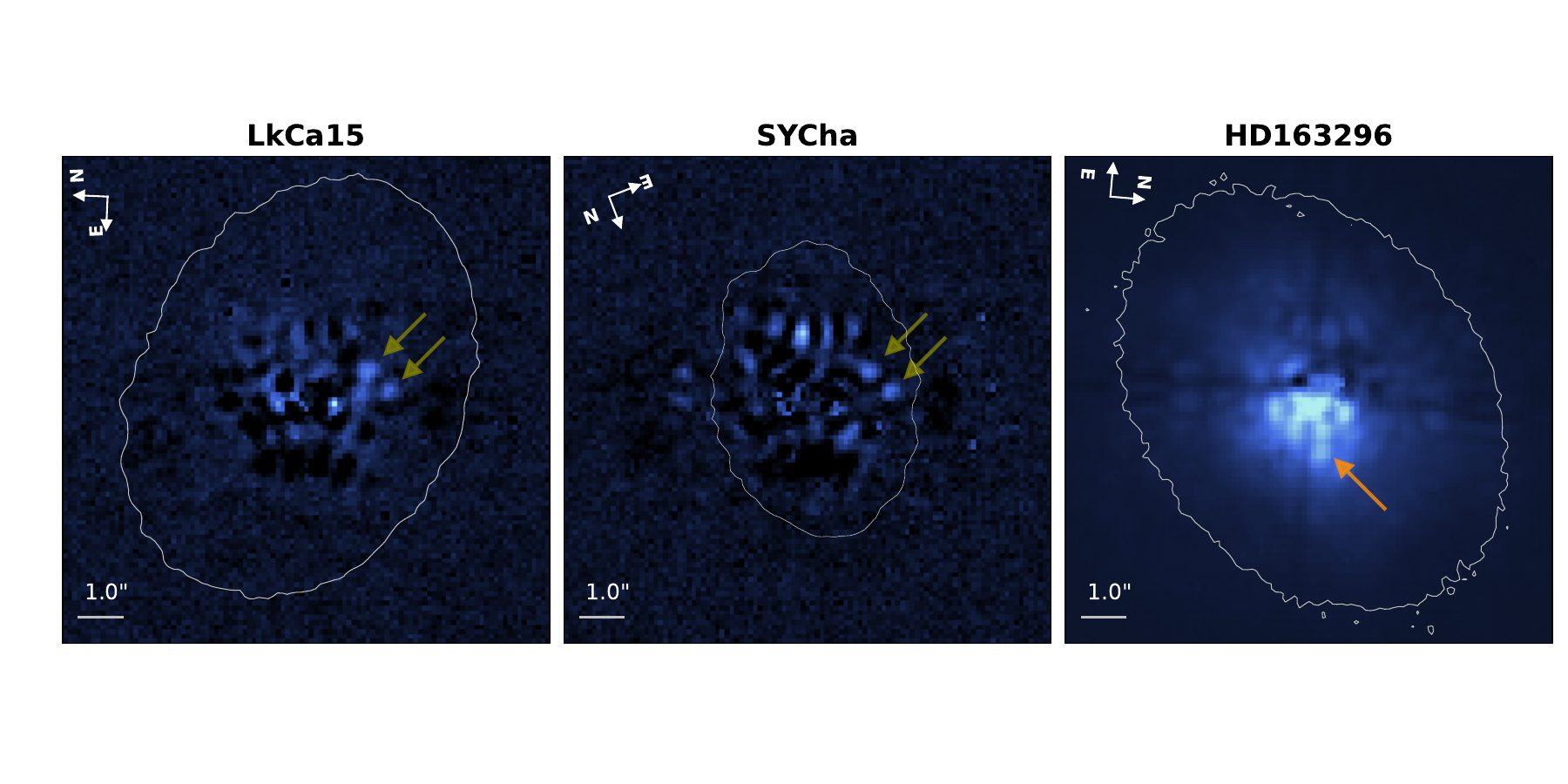}
    \caption{Residuals for LkCa15, SY~Cha and HD163296 (left to right) after the initial PSF subtraction but before being derotated to have North pointing upwards and East to the left. Multiple speckles at $1\farcs0-2\farcs0$ are consistent between different targets, especially as highlighted in the residual maps of LkCa15 and SY~Cha (green arrows). Other speckles appear unique in the presented sample, but they are also present in certain reductions of the other MIRI datasets in which a disk is present (e.g., speckle indicated by orange arrow in HD163296). \label{fig:speckles}}
\end{figure*}

\subsection{Speckles}
Besides our post-processing, the images show a number of speckles that could be interpreted as candidates. We highlight the major ones in Fig.~\ref{fig:speckles}, where we did not rotate the images to have North pointing up and East to the left. This allows us to compare the location of these speckles in the framework of the detector, comparing their position relative to the stellar diffraction. In particular, for HD163296 there is a bright signal highlighted with an orange arrow West of the star, and in both LkCa15 and SY Cha images, speckles remain in the residuals South and North-East of the star respectively (see green arrows). The potential signal in the residuals of HD163296 sits right on top of one of the speckles of the coronagraphic PSF, with a similar (albeit weaker) speckle present on the other side of the star. In addition we found speckles present at the same location in certain reductions of J1615 and AS209 in F1065C (PID7340, priv. comm.). 
Similarly, the two speckles in LkCa15 and SY~Cha (see Fig.~\ref{fig:speckles}) are along the coronagraphic arm and are co-located in the two sources when inspecting the images before derotating the data. The presence of multiple speckles colocated likely indicates that they are artifacts from diffraction and PSF subtraction rather than genuine sources. Given that these signals are not visible in other existing datasets of disk-less stars \citep[e.g.,][]{Boccaletti2024, Malin2024, Godoy2025, BendahanWest2026}, we conclude that they are likely caused by the presence of resolved emission and/or color mismatch between protoplanetary disk systems and reference stars. These artifacts are hard to subtract, as they are not reproduced in the `clean' reference star images.

\subsection{Planet detection limits}
\label{sec:detection_limits}
The noise distribution in ground-based high-contrast imaging data often deviates significantly from Gaussian statistics, invalidating the assumptions underlying standard detection metrics \citep[e.g.,][]{Bonse2023}. While a comparable analysis has not yet been carried out for JWST data, speckle-dominated regions of the residuals are likewise unlikely to follow a Gaussian distribution, especially when the diffraction of resolved inner disk signal makes PSF subtraction less effective. Moreover, the presence of extended emission from disks introduces spatial correlations between resolution elements, violating the independence assumption required for a standard t-test \citep{Bonse2023}.

We estimated detection limits by simply measuring signal-to-noise ratios (S/N) rather than confidence levels. Artificial point sources were injected into the data at varying flux levels until S/N=5 was reached following the methods described in \cite{Mawet2014}, in which the signal aperture ($r=1.7$~pix) is compared to the standard deviation of similar apertures at the same distance from the star correcting for small number statistics. The analysis was performed at radial separations spaced by 3.4 pixels ($0\farcs37$), corresponding to the FWHM of the MIRI PSF at 11.3~$\mu$m. At each separation, sources were injected at PAs $30^\circ$ and $60^\circ$ in each quadrant, while avoiding the coronagraphic support structures where throughput is strongly reduced. The limits in these areas are discussed further in Sect.~\ref{sec:discussion}. Synthetic point sources were generated using {\tt stpsf} at the appropriate detector locations and inserted into the pre-processed data. The PSF was then removed following the procedure described in Sect.~\ref{sec:methods}. For computational efficiency, the inner disk model was not refitted at each injection, but fixed to the model resulting from the median of the posterior values reported in Table~\ref{tab:results}. Flux densities measured at identical separations were averaged to derive the sensitivity at each radius. 

\begin{figure}[t]
    \centering
    \includegraphics[width=0.99\linewidth]{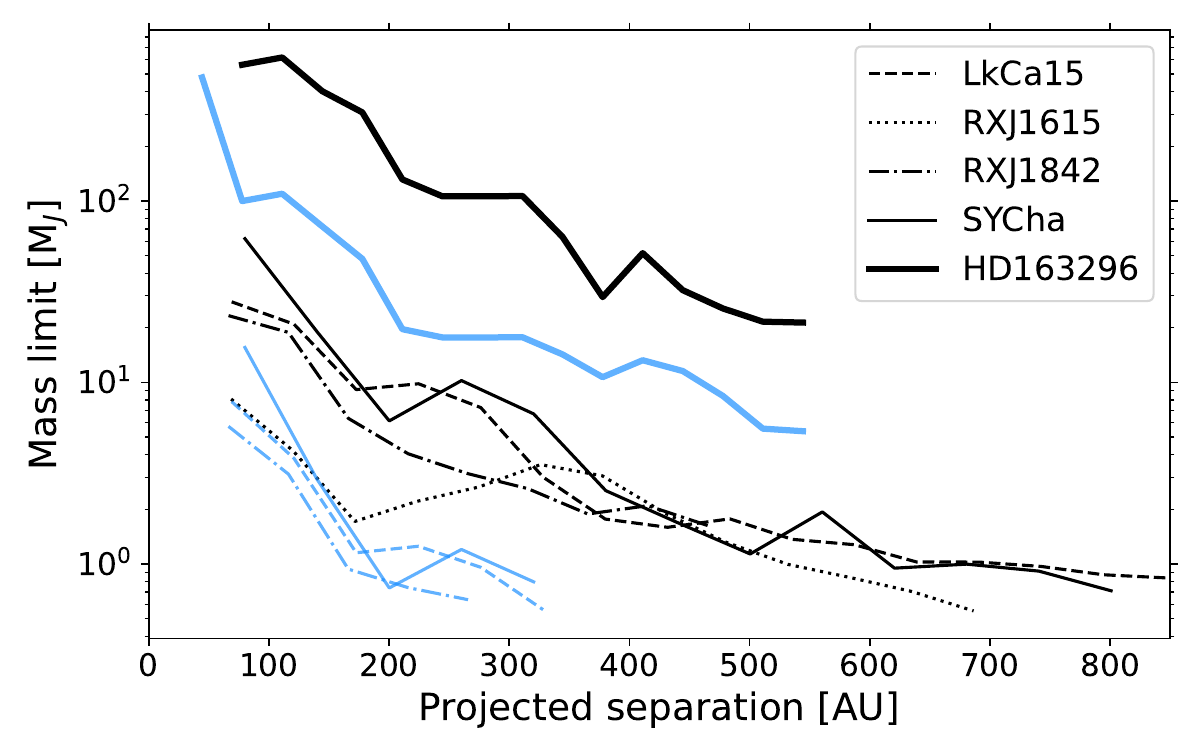}
    \caption{Mass sensitivity of our datasets as a function of the projected separation in au. Black lines assume the scenario in which the detectable flux is emitted from the planetary atmosphere, while the blue lines assume the same fractional CPD contribution as inferred for GQ~Lup~b, which has an estimated mass of $\sim30~\MJ$. Curves for the same target might have different radial extent if the evolutionary models did not cover certain flux ranges. In particular for RXJ1615, the mass limits when considering CPD contribution are deeper than the lowest mass reported in the ATMO models.}
    \label{fig:mass_limits}
\end{figure}

The resulting detection limits are shown in Fig.~\ref{fig:limits} for each target. In each panel, we report the limits 
after the initial PSF subtraction (Sect.~\ref{sec:initPSF}, solid blue line), 
after the removal of the inner disk (Sect.~\ref{sec:innerdisk}, dashed dark blue line), and for SY Cha and LkCa15 (the two targets without extended outer disk emission) after PSF subtraction with PCA (Sect.~\ref{sec:PCA}, dotted light blue line). 
The detection limits significantly improve in the inner $1\farcs5$ for RXJ1842 and LkCa 15 when the inner disk is removed. This is the area mostly impacted by the modeling and removal of the inner disk signal. Two systems show almost no improvement: HD163296 and J1615. These systems have bright emission from the outer disk that dominates the noise and is not removed with the modeling of the inner disk. Hence, the limits remain the same in both instances. For SY~Cha, removal of the inner disk does not significantly impact the limits, likely due to the irregular shape of the inner disk and its relatively faint flux. For SY Cha and LkCa15, the lack of extended emission from the outer disk allows us to perform PCA without resulting in strong oversubtraction. In that case the limits are better at separations larger than $\sim1\farcs0$, while the innermost region suffers from aggressive self-subtraction resulting in worse sensitivities.

\begin{figure*}[t]
    \centering
    \includegraphics[width=0.99\linewidth]{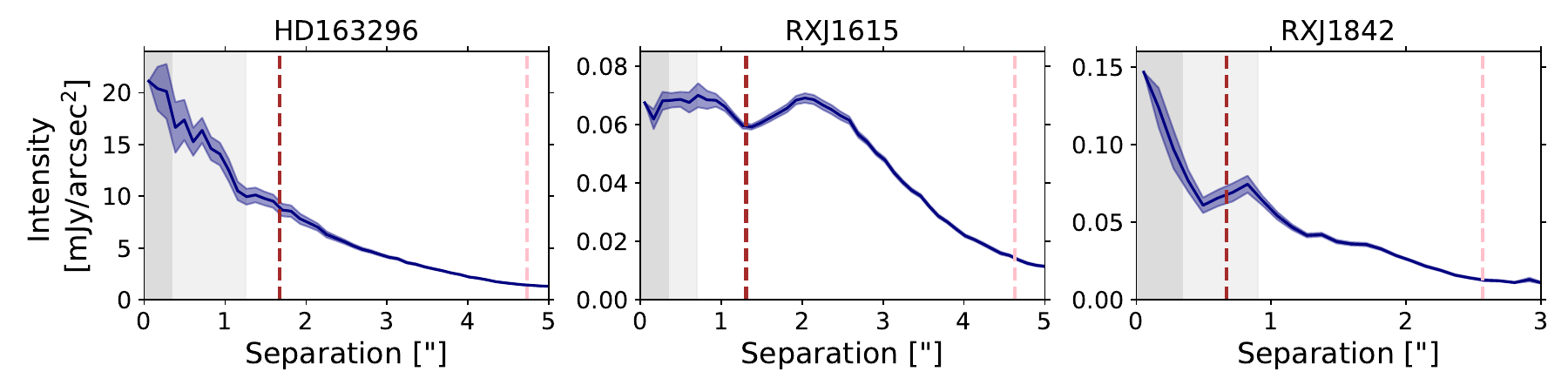}
    \caption{Intensity profile for the $11.3~\mu$m emission of the three disks detected in our sample. Dark blue profiles are extracted assuming the emission originates at the midplane and $^{12}$CO surface respectively. The grey area reports the inner working angle of MIRI at F1140C ($0\farcs34$), while the light grey area reports for each target the separation until which the residuals are dominated by speckles and therefore the profiles need to be taken with caution. Brown and pink vertical lines report the mm continuum and $^{12}$CO outer radii measured by the DSHARP, MAPS and exoALMA programs \citep{Huang2018, Law2021, Curone2025, GallowaySprietsma2025}. }
    \label{fig:profiles}
\end{figure*}

In Fig.~\ref{fig:limits} we also compare our detection limits with F1140C detections of different companions in the literature (see caption). In some cases our datasets reach the sensitivity to detect these companions, especially at larger separations, despite the additional systematics associated with the inner disk signal that is not present in those respective datasets. However, in the case of HD163296 none of the companions proposed so far are detectable, and the derived limits are considerably higher. This is due to a combination of two factors: (i) HD163296 is considerably brighter than most stars targeted with MIRI coronography ($F_{11.3~\mu m}=17.1$~Jy), and on average $\sim100\times$ brighter than the other targets in our sample; (ii) the outer disk emission is extremely bright up to the disk outer radius probed by $^{12}$CO at $\approx5\farcs0$, preventing the detection of anything fainter than $1-10$~mJy depending on the separation. Even a bright planetary mass companion with MIR excess emission from its circumplanetary disk like GQ Lup b \citep{Cugno2024} is far below the detection limits for HD163296. In the sixth panel of Fig.~\ref{fig:limits} we show again the LkCa15 limits, but this time we correct the flux measurements for the existing companions for the distance of LkCa15 (157 pc). Our data are not sensitive to any of those companions (with the exception of the planetary-mass companions GQ Lup b and Delorme 1 AB b) if they were placed at the distance of our objects.

We used the hot-start ATMO evolutionary models with chemical equilibrium \citep{Phillips2020} to convert the derived flux limits into masses assuming only atmospheric emission (black lines in Fig.~\ref{fig:mass_limits}). For each target, we used the best sensitivity provided at each separation by the curves in Fig.~\ref{fig:limits}. For HD163296, the sensitivities are so poor close to the star due to the bright disk that they reach a region of the parameter space not covered by the ATMO models. In those cases, potentially detectable objects enter the stellar regime, and we used the NextGen isochrones \citep{Allard1997} to assess mass detection limits. These mass limits highlight the limited capabilities of MIRI to detect forming planets in protoplanetary disks: only for J1615 (black dotted line) we have the sensitivity to detect $\sim2-3~\MJ$ planets at $\lesssim300$~au, approaching the mass range predicted by gas kinematics, while for the other sources we are only sensitive to planets in the $\sim3-20~\MJ$ range, and for separations $<100~$au our data could only detect planets with $M_p>10~\MJ$. These limits are at best comparable to the detection limits achieved with ground-based high-contrast imagers like SPHERE, NaCo and ERIS \citep{Asensio-Torres2021, Cugno2023}. HD163296 is clearly the least sensitive dataset, due to both the stellar brightness and the bright extended emission from  the outer disk. Even towards the edge of the disk around $\sim500~$au our sensitivity limits reach only $20~\MJ$. Although the achieved mass limits remain above those predicted from disk kinematics, these observations constitute the first attempt to search for kinematically inferred protoplanets at wavelengths where extinction is expected to be reduced, even if that reduction depends strongly on the grains properties \citep{Cugno2025}.

\section{Discussion}\label{sec:discussion}

\subsection{Inner and outer disk}\label{sec:disks}

\paragraph{Inner disk}
In protoplanetary disks, a significant fraction of the stellar radiation is intercepted by the disk and re-emitted in the mid-infrared. Radiative transfer models predict that this emission peaks at a few au and dominates the spectral energy distribution at $\sim11~\mu$m \citep[e.g.,][]{Mulders2012}. 
In addition, the light from the central star and from the inner disk is scattered by small grains in the surface layers and can also contribute significantly at these wavelengths \citep{Pinte2008}. Even when the emitting region is intrinsically compact ($\lesssim$ a few tens of au), diffraction from the coronagraphic mask redistributes this flux into extended structures that appear at separations of several arcseconds.
This effect significantly complicates PSF subtraction because the disk signal is not part of the stellar PSF and therefore cannot be reproduced by disk-free reference stars, causing the sensitivity of our data to be at least one order of magnitude lower than what can be achieved for a disk-free dataset like the one presented in \cite{Carter2023} as part of the JWST Early Research Science (ERS) Program (see Fig.~\ref{fig:limits}). Similar challenges have been reported in recent JWST/MIRI coronagraphic studies of systems hosting debris disks \citep{Boccaletti2024, Malin2024}, but since debris disks are intrinsically fainter, they result in significantly lower contamination. 

\paragraph{Outer disk} Three out of five disks, HD163296, J1615 and J1842, also show extended emission at $11.3~\mu$m clearly detected in the residuals (Fig.~\ref{fig:gallery}), out to the outer disk radius as inferred from the $^{12}$CO line emission. 
Figure~\ref{fig:profiles} reports the intensity profiles extracted using the {\tt goFish} package \citep{Teague2019} for these three disks. {\tt goFish} is designed to extract radial profiles from spatially resolved disk observations by accounting for the projected disk geometry and, when available, for an elevated emitting surface. This is particularly useful for comparing tracers that may originate above the disk midplane, as expected for optically thick CO emission and for scattered-light surfaces. As we do not know the exact disk surface heights that contribute to the $11.3~\mu$m signal, the profiles were extracted assuming emitting surface shapes inferred from $^{12}$CO line emission \citep{Law2021_V, GallowaySprietsma2025}. This choice is motivated by previous work showing that the CO emitting surface can lie at heights comparable to those traced by near-infrared scattered light \citep{Rich2021}, making it a reasonable approximation for emission emerging from the disk surface.
Overplotted are the inner working angle of the MIRI data (dark grey area) and the regions where residual speckles could still impact the measured signal based on the residuals in Fig.~\ref{fig:gallery} (light grey area), the outer radius of the continuum mm emission and of the $^{12}$CO line emission. In all three cases, the profiles confirm that MIRI detects disk signal across the entire extent of disk (as defined by the CO emission). We find a clear intensity drop in J1615, co-located with the outer radius of the continuum emission. The intensity drop in J1842 at 0.5\arcsec{}, with the subsequent brighter ring, is due to PSF subtraction and inner disk modeling artifacts. Indeed, the radial location of these features match some of the residuals visible in Fig.~\ref{fig:gallery}, in particular some bright ring-like residuals from the diffraction pattern on the West side of the star. Both J1842 and HD163296 have a smooth outer disk signal.  

This extended signal could trace stellar light scattered off by the small dust grains in the surface. This has been seen at shorter wavelengths for HD163296  
in the HST/STIS image of the system  that shows disk signal extending to $\sim$3.6\arcsec{} in radius \citep{Grady2000,Rich2020, Benisty2023}. Although the overall contribution of scattered light from small dust grains decreases at longer wavelengths, it could still produce a signal detectable from space with MIRI. A related interpretation has recently emerged from JWST observations of edge-on disks, where spatially extended mid-infrared continuum and molecular-line emission can be explained as inner-disk radiation scattered by dust grains located high in the outer disk surface. In some cases, reproducing the MIRI observations appears to require grains with sizes of order several to tens of microns at high altitude \citep{Duchene2024, Villenave2024}, which is not straightforwardly expected from simple dust-settling calculations

The alternative explanation for the bright extended disk signal detected in our MIRI residuals is the emission from Polycyclic-Aromatic Hydrocarbon (PAH) emission. Indeed, the 11.3~$\mu$m PAH feature is present in the spectral region covered by the F1140C filter. This possibility is particularly relevant in light of recent JWST studies of edge-on disks \citep[e.g.,][]{Arulanantham2024, Bergner2026}, which show that PAH emission can be spatially extended across disk surfaces. In edge-on systems, the 11.3~$\mu$m PAH emission can be co-spatial with the scattered-light continuum and may extend over large disk scales.
In HD163296, long-slit spectroscopy observations with VISIR-NEAR spatially resolved PAH emission in the 7.9 and 8.6 µm bands \citep{Yoffe2023}. The 11.3~$\mu$m feature is however degenerate with crystalline forsterite emission arising from the inner disk region, and the PAH contribution is thus unclear. 
For T Tauri stars, the presence of PAHs in their disks is uncertain as the stars have lower UV emission, and thus less PAH excitation \citep{Siebenmorgen2010, Lange2025}. Futhermore, PAH clustering may decrease the PAH abundance in the upper layers of T Tauri disks \citep{Lange2021}. In practice, J1615 does not exhibit clear PAH lines from the MRS spectra \citep{Volz2026}. Those two possibilities will be further discussed in Stapper et al. in prep. 

If PAH emission contributes significantly to the extended F1140C signal, the choice of this filter has important consequences for future programs. F1140C can be advantageous for studies aiming to trace PAH-rich disks or extended mid-infrared disk emission, but it may be suboptimal for planet searches in disks where spatially structured PAH emission increases the background and complicates PSF subtraction. If this is the case, filters less affected by strong PAH bands may provide cleaner contrast for searches focused on compact protoplanet emission.

\subsection{Detection limits}
\paragraph{Comparison with kinematics} As the original goal of the program is to detect embedded planets in 5 disks for which kinematical signatures were detected, we compare those predictions to our detection limits. For the candidates protoplanets from the exoALMA program, \citet{Pinte2025} report the planetary locations adopted in their models to reproduce the kinematic signatures seen in channel maps. These separations range from $\sim$0.6\arcsec{} to 1.9\arcsec{}. The dynamically inferred masses of these candidates lie between 1 and 5 $\MJ$, whereas our detection limits make us sensitive only to planets in the $\sim$3–20 $\MJ$ range for the exoALMA sources when assuming the ATMO evolutionary models, where the $3-5~\MJ$ regime is hit only in RXJ1615 at separations lower than the bright ring and in RXJ1842. 

\begin{figure*}[t]
    \centering
    \includegraphics[width=0.99\linewidth]{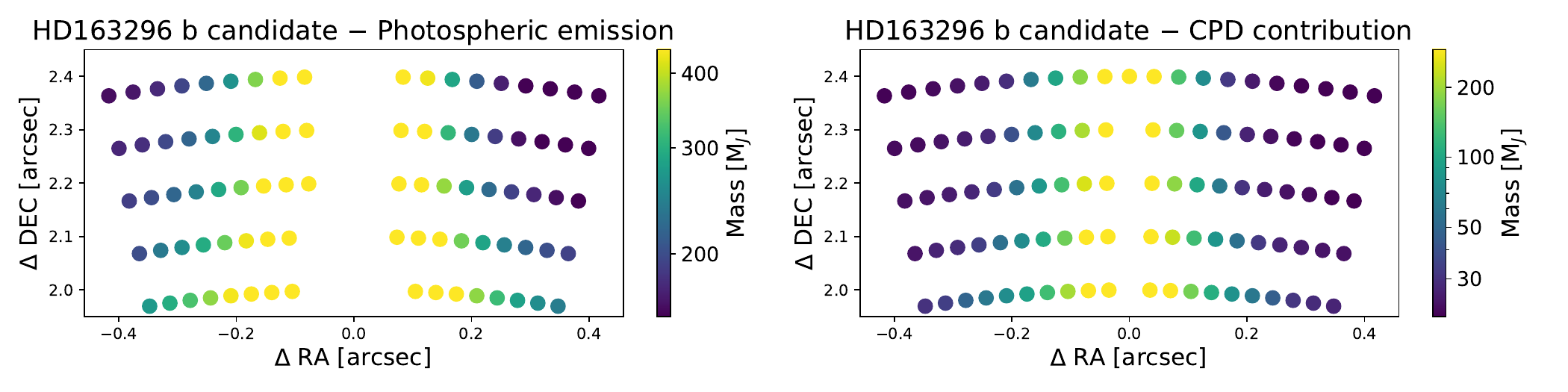}
    \caption{Detection mass limits as a function of separation and PA for the candidate protoplanet HD163296b. Left panel shows the case in which all the emission is atmospheric, while the right panel reports the case in which only 14\% of the emission comes from the atmosphere, while the rest comes from emission produced by circumplanetary environment. This value was extracted for the GQ Lup b planetary mass companion, which is hosting a CPD detected in the MIR \citep{Cugno2024}. }
    \label{fig:HD163296}
\end{figure*}

In the case of HD163296, two planet candidates were inferred from the gas kinematics 
\citep{Teague2018a, Pinte2018b,Izquierdo2022, Izquierdo2026}. For the outermost one, the current predictions place the planet at a separation of $2\farcs3\pm0\farcs2$ and position angle $0\pm10^\circ$. This region of the detector is crossed by the transition of the FQPM coronagraph, where the throughput can be extremely low for an extension of several pixels, and because of the position of HD163296 in the sky, JWST/MIRI can only observe it in this orientation. Thus, we explore this area of the detector by inserting artificial signals in steps of $0\farcs1$ and $1^\circ$ and determining the local detection limits. The resulting mass limits are reported in the left panel of Fig.~\ref{fig:HD163296}, clearly showing that the low coronagraph throughput harms our ability to search for low-mass forming companions (compare with Fig.~\ref{fig:mass_limits}). Unfortunately, a different spacecraft orientation that enables a different positioning of the FQPM was not possible due to HD163296's coordinates in the sky. 
At the predicted location of the inner planet (0.75\arcsec{}), the throughput is only $\sim0.3$ and we are only able to exclude objects with a flux density $>64.48$~mJy. When considering only atmospheric emission, this value translates into an object with $1.5~M_\odot$.

We note that \cite{Uyama2025} reported JWST/NIRCam observations of HD163296 in the F410M filter. Assuming no extinction ($A_V=0$ mag), they reach a sensitivity of $1-2~\MJ$ for hot/warm-start evolutionary models at the location of the outer candidate. This limit is significantly lower than what we obtain from the MIRI data, but a direct comparison remains difficult. We abstain from such a comparison because their ADI sequence enables a more effective removal of the extended disk emission, whereas the extremely bright disk residuals in our MIRI data dominate the background and significantly degrade the sensitivity limits. Obtaining both NIRCam and MIRI observations for a target that does not suffer from these systematics would enable a more reliable comparison, and would help determine which instrument, PSF subtraction strategy and wavelength regime are best suited for these searches in the future.

\paragraph{Emission from circumplanetary material}
The emission from the circumplanetary environment could significantly contribute at mid-IR wavelengths \citep{Zhu2015}. Understanding the relative contribution of this circumplanetary environment and planetary atmospheres is critical for interpreting our detection limits. \citet{Blakely2025} modeled the available spectra and photometry of the PDS\,70 protoplanets under scenarios including and excluding MIR emission from a CPD. For their best-fit solutions, the predicted flux in the F1140C band differs by nearly two orders of magnitude depending on whether a CPD is included. Such a large factor would imply that CPD thermal emission can dominate over atmospheric emission at $\sim10~\mu$m. While this highlights the diagnostic power of mid-infrared observations, a contrast of this magnitude remains uncertain and widely untested. Additional constraints with instruments such as METIS on the ELT will be required to establish robust CPD-to-atmosphere flux ratios for embedded protoplanets. 

Another potentially informative system is WISPIT2bc, a young system with an accreting protoplanet detected at $0\farcs33$ from its host star and a second companion within the central cavity which seems not to be accreting anymore \citep{van_Capelleveen2025, Close2025, Lawlor2026}. Although ongoing accretion is confirmed through H$\alpha$ emission for b, no clear mid-infrared excess attributable to a CPD has been reported, and high-resolution millimeter continuum observations yielded a non-detection for a potential CPD \citep{Facchini2026}. Thus, currently WISPIT2~b does not provide quantitative constraints on the mid-IR CPD-to-atmosphere flux ratio.

An alternative approach is to consider more massive planetary-mass objects (PMOs) with published MIRI spectra. Two such objects are Delorme~1~AB~b \citep{Malin2025_delorme} and GQ~Lup~b \citep{Cugno2024}. Delorme~1~AB~b is significantly older ($\sim$40 Myr) and could therefore represent a different evolutionary stage, however GQ~Lup~b is much younger ($2-5$~Myr, \citealt{MacGregor2017}) and more representative of the evolutionary stage of embedded protoplanets. Using the MIRI MRS spectrum presented by \citet{Cugno2024}, we estimate that only $\sim14\%$ of the flux within the F1140C bandpass originates from atmospheric emission, with the remaining $\sim86\%$ attributable to circumplanetary material. Although GQ~Lup~b is relatively massive ($\sim30~M_{\rm Jup}$; \citealt{Stolker2021}), it provides an empirical indication of how much a CPD could contribute in the mid-infrared for young low-mass companions.

The right panel of Fig.~\ref{fig:HD163296} illustrates the case in which 86\% of the total F1140C flux for the HD163296b protoplanet arises from circumplanetary material, analogous to GQ~Lup~b. Under this assumption, our observations probe substantially lower companion masses than if only atmospheric emission were present. Nevertheless, we remain above the canonical planetary-mass regime and significantly above the $\sim2~M_{\rm Jup}$ mass inferred from disk kinematics \citep{Teague2018a, Pinte2018b}. The comparison between the top and bottom panels of Fig.~\ref{fig:HD163296} demonstrates the strong leverage provided by mid-IR observations when circumplanetary emission contributes significantly at long wavelengths.


Adopting the same fractional CPD contribution as inferred for GQ~Lup~b, the flux density limits derived for the planet candidate identified at 94 au \citep{Izquierdo2026} exclude companions more massive than $\sim0.5~M_\odot$. However, these limits remain insufficient to probe the mass range predicted for the kinematic protoplanet candidate ($\lesssim1~M_{\rm Jup}$; \citealt{Izquierdo2026}).

For the outermost planet, the absence of a second spacecraft roll likely limits our sensitivity. A second roll would both recover regions currently affected by reduced coronagraphic throughput and improve PSF subtraction via angular differential imaging without risking the self-subtraction of a potential companion signal at these separations. Such observing strategies are likely essential for fully exploiting the diagnostic potential of mid-infrared coronagraphy for embedded protoplanets and should be explored in future programs. However, for the inner planet, the addition of a second spacecraft roll would likely provide only marginal improvement as the significant spatial overlap of the protoplanet signal in the two rolls at only $0\farcs75$ would result in substantial self-subtraction during PSF removal. Progress toward deeper limits requires improved forward modeling of both the inner disk diffraction pattern and the large-scale disk emission.




\section{Summary and Conclusions}\label{sec:conclusion}

We presented the first JWST/MIRI coronagraphic observations of five protoplanetary disks. Our targets host candidate embedded planets inferred from ALMA gas kinematics at large separations ($\gtrsim70$~au). Our goal was to directly detect the companions responsible for the observed velocity perturbations. The main results are: 
\begin{enumerate}[label=\roman*)]
    \item No point source consistent with the kinematically inferred planets is detected in the five systems. Our observations are typically sensitive to companions of $\gtrsim3–20~\MJ$ at separations of a few hundred au, only marginally overlapping with the predictions from kinematic signatures, which suggest planets of only $\sim1–5~\MJ$. Assuming significant contribution from circumplanetary material, these sensitivity are significantly improved, but it is not clear to what extent the circumplanetary environment can contribute.
    \item The residuals show persistent speckles that appear at similar detector locations across multiple datasets, highlighting the need for careful validation of candidates in protoplanetary disks. Moving forward, when candidate protoplanets are investigated, it will be good practice to cross-check the positions with the location of known residuals patterns for these types of objects. 
    \item In HD163296, J1615, and J1842 we detect extended emission tracing the full radial extent of the disks, likely produced by scattered light and/or PAH emission within the F1140C bandpass, enabling a characterization of their mid-IR outer disk emission.  
    \item Bright inner disks produce diffraction patterns that propagate to arcsecond scales and significantly reduce the achievable contrast compared to disk-free systems. We present a new method to characterize, reproduce and subtract this inner disk contribution from the images, which can be improved through better modeling of the instrumental PSF. These new mid-IR constraints can provide useful insights to better understand the disk grain properties and distribution at short separations. 
    \item It is still unclear what is the best JWST instrument to search for protoplanets, and the results likely depend on the presence of bright disk emission at $<10~\mu$m, the separation of the companion (which enables or prevent the use of ADI), and the presence of warm circumplanetary material. 
\end{enumerate}

These results highlight the fundamental challenge of directly imaging forming protoplanets embedded in bright protoplanetary disks with JWST/MIRI. Optimized observing strategies and improved instrumental modeling of the diffraction patterns of resolved sources will be key to improve these sensitivities moving forward. Looking ahead, although intrinsic disk-related challenges will likely persist in the absence of a clearly depleted gap or cavity, such as those in PDS 70 or WISPIT2, the next generation of high-contrast mid-infrared instruments, in particular METIS on the ELT \citep{Brandl2014}, will provide higher angular resolution and sensitivity, enabling the detection and characterization of forming planets and their circumplanetary environments at separations inaccessible by JWST.

\begin{acknowledgments}
GC thanks the Swiss National Science Foundation for financial support under grant numbers P500PT\_206785 and P5R5PT\_225479. This work has been carried out within the framework of the NCCR PlanetS supported by the Swiss National Science Foundation under grant 51NF40\_205606. MB and LMS have received funding from the European Research Council (ERC) under the European Union’s Horizon 2020 research and innovation programme (PROTOPLANETS, grant agreement No. 101002188). 

Support for C.J.L. was provided by NASA through the NASA Hubble Fellowship grant No. HST-HF2-51535.001-A awarded by the Space Telescope Science Institute, which is operated by the Association of Universities for Research in Astronomy, Inc., for NASA, under contract NAS5-26555.
CHR acknowledges the support of the Deutsche Forschungsgemeinschaft (DFG, German Research Foundation) Research Unit "Transition discs" - 325594231. CHR is grateful for support from the Max Planck Society.
GG received funding from the European Research Council (ERC) under the European Union’s Horizon Europe research and innovation program (grant agreement No. 101053020, project Dust2Planets). MB-A received funding from the European Union's Horizon Europe research and innovation programme under the Marie Skłodowska-Curie grant agreement No. 101281707 (InnerDisk). Views and opinions expressed are however those of the author(s) only and do not necessarily reflect those of the European Union or the European Research Executive Agency. Neither the European Union nor the granting authority can be held responsible for them.
    
This work is based on observations made with the NASA/ESA/CSA James Webb Space Telescope. The data were obtained from the Mikulski Archive for Space Telescopes at the Space Telescope Science Institute, which is operated by the Association of Universities for Research in Astronomy, Inc., under NASA contract NAS 5-03127 for JWST. These observations are associated with programs GO 2153 (Grant \#JWST-GO-02153.002-A) and GO 3254 (Grant \#JWST-GO-03254.002-A). The specific observations analyzed in this work can be accessed via \url{ https://doi.org/10.17909/ycgj-nh17}.
This research has made use of the Jean-Marie Mariotti Center \texttt{SearchCal} service, which involves the JSDC and JMDC catalogues.
\end{acknowledgments}

\begin{contribution}
GC and MB carried out the data reduction and analysis, and wrote the manuscript. GC developed the inner disk forward modeling algorithm, with contributions from AB, MP and MM. GC, PP, MB and LP prepared the observations. GC, MB, RT, SF, MF and CP led the observing proposals. 
All other authors read and provided feedback on the manuscript. 


\end{contribution}

%
\facility{JWST(MIRI)}

\software{astropy \citep{2013A&A...558A..33A,2018AJ....156..123A,2022ApJ...935..167A},  
          stpsf \citep{Perrin2014},
          spaceklip (version 2.7.1) \citep{Kammerer2022, Carter2023},
          species \citep{Stolker2020_miracles}, 
          GoFish \citep{Teague_gofish}, 
          jwst \citep{jwst_pip}, 
          pymultinest \citep{Feroz2009, Buchner2016}, 
          }


\appendix
\restartappendixnumbering

\bibliography{biblio}{}
\bibliographystyle{aasjournalv7}



\end{document}